\documentclass[twoside]{elsart}
\usepackage{epsfig}
\usepackage{psfig,wrapfig,amsmath,amstext}

\newcommand{\beq}{\begin{equation}}
\newcommand{\eeq}{\end{equation}}
\newcommand{\beqx}{\begin{displaymath}}
\newcommand{\eeqx}{\end{displaymath}}
\newcommand{\beqar}{\begin{eqnarray}}
\newcommand{\eeqar}{\end{eqnarray}}
\def\gev{\,\mbox{GeV}}

\begin{document}
{
\null \hfill \begin{minipage}{5.5cm} Freiburg-THEP-06/05\\
~~~~~~\\
{\it Where~is~Heidi~hiding~?}\\
{\it Heidi~is~hidden}\\
{\it in~the~high-D~Higgs~Hill~!}\\
\end{minipage}
}
\begin{center}
\title{A higher dimensional explanation of the excess of 
 Higgs-like events at CERN LEP}
\author{J.~J.~van~der~Bij and S.~Dilcher}\\
\address{Institut f\"ur Physik, Universit\"at Freiburg,\\
  Hermann-Herder-Str. 3, 79104 Freiburg
i. Br., Germany}
\end{center}
{\bf Abstract}\\
Searches for the SM Higgs boson by the four LEP experiments have found
a $2.3\,\sigma$ excess at $98\,\mbox{GeV}$ and a smaller $1.7\,\sigma$ at around
$115\,\mbox{GeV}$. We interpret these excesses
as evidence for a Higgs boson coupled to a higher
dimensional singlet scalar. The fit implies a relatively low dimensional
mixing scale $\mu_{lhd} < 50\gev$, which explains the low confidence level
found for the background fit in the range $s^{1/2} > 100\gev$. The
data show a slight preference for a five-dimensional over a  six-dimensional field.
This Higgs boson cannot be seen at the LHC, but can be studied at the ILC.


With the latest developments from high energy colliders like LEP
and the Tevatron the standard model (SM) has been established up to
the loop level.  The main missing ingredient is the direct detection of
the SM Higgs boson. The four LEP experiments ALEPH, DELPHI, L3 and OPAL
have extensively searched for the Higgs boson. The final combined result
has been published in \cite{lephiggs}. The absence of a clear signal has led to
a lower limit on the Higgs boson mass of 114.4 GeV at the 95\% confidence level.
Although no clear signal was found the data have some intriguing features,
that can be interpreted as evidence for Higgs bosons beyond the standard
model. There is a $2.3\,\sigma$ effect seen by all experiments at around 98 GeV.
A somewhat less significant $1.7\,\sigma$ excess is seen around 115 GeV. Finally
over the whole range $s^{1/2} > 100\gev$ the confidence level is less than
expected from background.\\

 Within the minimal supersymmetric standard model (MSSM)
and other extensions [2-5], the excesses at 98 GeV and 115 GeV were interpreted as
evidence for the presence of two Higgs bosons $H_i$ with couplings reduced
by a factor $\alpha_i$ to matter
$g^2_{i} = \alpha_i\, g^2_{SM-Higgs}$. We will call such Higgs bosons fractional Higgses
in the following. The excess at 98 GeV is well described by a 10\% fractional Higgs.
More precisely \cite{drees} gives limits $0.056 < \alpha_{98} < 0.114$ and
a mass range $95\gev < m_{Higgs} < 101\gev$. The second peak at 115 GeV is then
interpreted as a second Higgs boson with $\alpha_{115} = 0.9$. The first peak
at 98 GeV is rather convincing. The second one at 115 GeV is  compatible with
the data, but not really preferred as the
data at 115 GeV are also compatible with pure background with a similar confidence level.
The data correspond roughly to background plus one half of a Higgs boson,
 however with a large 
uncertainty. The MSSM fit is therefore not completely satisfactory and it is natural
to look for other extensions to describe the data.\\  

Actually the precision measurements leave only very little
space for extensions, as these tend to spoil the agreement with experiment
due to a variety of effects, one of the most important of which is
the appearance of flavor-changing neutral currents. Even the MSSM has to finely
tune a number of parameters. This leaves only one type of extensions that are
safe, namely the
singlet extensions. Experimentally  right handed
neutrino's appear to exist. Since these are singlets a natural extension of the SM
is the existence of singlet scalars too. These will only have a very limited effect
on radiative corrections, since they appear only in two-loop calculations.
For a mini-review on singlet extensions see \cite{mnmsm}.\\

In this letter we will try to fit the data within the specific model of
a SM Higgs boson coupled to a higher dimensional ($d=5, 6$) singlet scalar,
where we  assume that the higher dimensions are open and flat.
This leads to a Higgs propagator of the form:

\beq
D_{H\,H}(q^2)= \left[ q^2 +M^2 - \mu_{lhd}^{8-d}
(q^2+m^2)^{d-6 \over 2} \right]^{-1}. \eeq

The masses $M$, $m$ and $\mu_{lhd}$ are the free parameters of the model and
are defined more precisely later. The scale $\mu_{lhd}$ stands for
low-to-high-dimensional mixing mass and measures the coupling of the high-$d$
singlet scalar and the 4-$d$ doublet.
The propagator contains a particle peak and a continuum.
This model has, like the MSSM,  three free parameters to fit the data.
We will take the excess at 98 GeV at face value and interpret it as the delta-peak
in the propagator. The excess at 115 GeV is interpreted as an enhancement
due to the continuum of the Higgs propagator. Because of the uncertainty of this
excess we will only demand that a Higgs integrated spectral
density $\int \rho(s) ds > 30\%$
is present in the range $110\gev < s^{1/2} < 120\gev$. We will find that
with these conditions one gets a relatively small highdimension-lowdimension
mixing scale $\mu_{lhd} < 50\gev$. This result has a non-trivial consequence.
It implies that
the continuum starts very close to the delta-peak. This would therefore explain 
why in the range $s^{1/2} > 100\gev$ the confidence level of the data is less
than expected for background. This cannot be explained by the MSSM.\\

The effects of singlets appear in two forms, one is the mixing with the SM
Higgs, the other is invisible decay. In this paper we are
interested in pure mixing models.  It is actually possible to have a Higgs model
that has only Higgs-mixing. If one starts with an interaction of the form
$H \Phi^{\dagger}\Phi $, where H is the new singlet Higgs field and $\Phi$ the SM
Higgs field, no interaction of the form $H^3$ or $H^4$ is generated with an infinite
coefficient \cite{hill}. Therefore one can leave the $H^3$ and $H^4$ interactions
out of the Lagrangian without violating renormalizability. This type of model can easily
be extended to more fields $H_i$.
The situation becomes quite interesting, when one allows for an infinite number
of fields, in particular when one assumes the field $H$ to be moving in 
$d=4+\delta$ dimensions.
Normally speaking this would lead to a nonrenormalizable theory. However
since the only interaction is of the form
$H\Phi^{\dagger}\Phi$, which is superrenormalizable in four dimensions,
the theory stays renormalizable. An analysis of the power counting of divergences
shows that one can associate the canonical dimension $1+\delta/2$ to the
$H$-field. This means that the theory stays renormalizable as long as
$\delta \le 2$. 
When one assumes, that the extra dimensions are compact, for instance a torus,
with radius $R=L/2 \pi$,
one can simply  expand the $d$-dimensional field
$H(x)$ in terms of Fourier modes:
\begin{equation}
H (x) ={1 \over \sqrt{2}\, L^{\delta/2}} \sum_{\vec{k}} H_{\vec{k}}(x_\mu)\,
e^{i {2 \pi \over L} \vec{k} \vec{x} },  \;\;\;\;\;\;\;\;\;\;\;\;
  H_{\vec{k}}      =   H_{-\vec{k}}^*~~~.        
\end{equation}

Here $x_\mu$ is a four-vector, $\vec{x}$ is $\delta$-dimensional
and the $\delta$ components of $\vec{k}$ take only integer values.

One then finds a Lagrangian of the form:
 \begin{equation}
  L=-\frac{1}{2}D_{\mu} \Phi^{\dagger} D_{\mu} \Phi 
 - {1 \over 2} \sum (\partial_{\mu} H_k )^2
 +{M_0^2 \over 4} \Phi^{\dagger}\Phi 
-{\lambda \over 8} (\Phi^{\dagger}\Phi)^2
 -\sum {m_k^2\over2} H_k^2 
 -{g \over 2} \Phi^{\dagger}\Phi  \sum H_k~. 
\end{equation}
The masses of the $H_k$ fields are given by $m_k^2=m_0^2+m_d^2\vec k^2$.
Here $\vec k$ is a $\delta\mbox{-dimensional}$ vector and $m_0$ is a $d$-dimensional
mass term for the field $H$.  This term is necessary to insure stability
of the vacuum. The last term can be written as a so-called brane-bulk term
\begin{equation} 
 S=\int d^{4+\delta}x \prod_{i=1}^{\delta} \delta(x_{4+i}) H(x) 
\Phi^{\dagger}\Phi .   
\end{equation}
This Lagrangian leads to spontaneous symmetry breaking. Both the
$\Phi$-field and the $H_k$-fields will receive a vacuum expectation value,
which we call $v$ respectively $h_k$. In the unitary gauge we write
$\Phi=v+\phi$ and $H_k=h_k+x_k$.

The conditions for the minimum of the potential $V'=0$ read:
\beq {\partial V \over \partial \Phi}|_{v,h_k}=
{\lambda \over 2} v^3 - {M_0^2 \over 2} v
+g v \sum_k h_k =0 ,\eeq
\beq {\partial V \over \partial H_k}|_{v,h_k}=m_k^2 h_k + {g v^2 \over 2} =0 ,\eeq
\beq \Rightarrow v^2={M_0^2 \over \lambda-g^2 \sum_k {1 \over m_k^2}} 
\;\; ; \;\; h_k={-g M_0^2 \over 2 m_k^2 (\lambda-g^2 \sum_k {1 \over m_k^2}) } .\eeq
From the second derivatives one then finds the mass matrix  :
\beq { L}_{mass}={-M^2 \over 2} \phi^2 -\sum_k {m_k^2 \over 2}x_k^2 +gv 
\phi \sum_k  x_k ,\eeq
 with  $ M^2={3 \over 2}\lambda v^2 - {M_0^2 \over2 } +g \sum_k h_k = \lambda v^2 $.\\

One can derive the Higgs propagator without knowing the eigenstates
explicitely by inverting the matrix:

\beq D_{ij}(q^2)^{-1}=  \left( \begin{array}{ccc}
q^2+M^2 & -gv & \dots \\
-gv & q^2+m_i^2 & 0  \\ 
\vdots & 0 & \ddots \\
\end{array} \right)  .\eeq

First we define the auxiliary quantity $Q$:
\beq
Q = 1-\sum_k {g^2 v^2 \over (q^2+M^2) (q^2+m_k^2)} .
\eeq
 $D_{ij}(q^2)$ is found to be:

\beq 
D_{ij}(q^2)=
 \frac{1}{Q}
 \left( \begin{array}{cc}
\frac{1}{q^2+M^2} & 
~~~~~~\frac{gv}{ (q^2+M^2) (q^2+m_j^2) }\\
\frac{gv} { (q^2+M^2) (q^2+m_i^2) }  & 
~~~\left( \frac{Q} {q^2+m_i^2} \delta_{ij} +
\frac{g^2 v^2}{(q^2+M^2)(q^2+m_i^2)(q^2+m_j^2)}  \right)\\ 

\end{array} \right) . \eeq

In the continuous case the Higgs propagator, which corresponds to the (1,1)
component of D, becomes:
\beq
D_{H\,H}(q^2)= \left[ q^2 +M^2 -g^2 v^2 {\Gamma (1-\delta/2) \over (4 \pi)^{\delta/2} }
(q^2+m_0^2)^{\delta-2 \over 2} \right]^{-1} .\eeq

After redefining variables this becomes:

\beq
D_{H\,H}(q^2)= \left[ q^2 +M^2 - \mu_{lhd}^{8-d}
(q^2+m^2)^{d-6 \over 2} \right]^{-1} .\eeq

We will now consider in more detail the
two cases $d=5$ and $d=6$ separately and compare these with
the data. Other values of $d$ are in principle possible, but have no simple physical
interpretation as higher dimensional fields.\\

First we consider the case $d=5$. 
The propagator becomes of the
form:
\beq
D_{H\,H}(q^2)= \left[ q^2 +M^2 - \mu_{lhd}^3
(q^2+m^2)^{-1/2} \right]^{-1} .\eeq

This corresponds to a K\"all\'en-Lehmann spectral density:
\begin{eqnarray}
\rho(s) = &\theta(m^2-s)\,\, \frac{2(m^2-s_{peak})^{3/2}}{2(m^2-s_{peak})^{3/2}
+\mu_{lhd}^3}
\,\,\delta(s-s_{peak}) \nonumber\\   
+& \frac{\theta(s-m^2)}{\pi}\,\,
\frac{\mu_{lhd}^3\,(s-m^2)^{1/2}}{(s-m^2)(s-M^2)^2+\mu_{lhd}^6} ,
\end{eqnarray}
where $s_{peak}$ satisfies:
\beq
s_{peak}-M^2+\mu_{lhd}^3(m^2-s_{peak})^{-1/2} = 0 .\eeq

The peak  always exists at a positive value of $s_{peak}$ as long as
$\mu_{lhd}^3 < m M^2$, which we assume satisfied, since this corresponds
to the condition that we expand around a minimum of the potential.
Actually this form of the spectral density is only correct as long
as $M>m$. In the case $m>M$ in principle more peaks can appear,
however this would imply for the residue at the pole $\alpha_{98} > 1/3$,
which is not allowed by the data.

We now consider the case $d=6$.
This case is special, since it corresponds to the limiting dimension, where
the theory is still renormalizable.
In the limit $d \rightarrow 6$ one notices that the propagator
has  a logarithmic singularity. This is because the coupling constant
becomes dimensionless and is running as a function of the renormalization scale.
In contrast to normal four-dimensional models, the running appears
already at the tree level of the theory.
Specializing to this case and expanding the $\Gamma$-function
around $d=6$ the propagator can be written as:

\beq
D_{H\,H}(q^2)= \left[ q^2 +M^2 +\mu_{lhd}^2\,
\log(\frac{q^2+m^2}{\mu_{lhd}^2}) \right]^{-1} .
\eeq

The spectrum contains a pole term for a positive mass as long
as:
\beq M^2 +\mu_{lhd}^2 \log(m^2/\mu_{lhd}^2) > 0 .\eeq
When this condition is not fulfilled the pole becomes a tachyon.
The reason is as before in the $d=5$ case,
that in this case the potential due to 
the attractive $H$-exchange dominates over the repulsive
$\lambda \phi^4$ term and the vacuum is not stable.

The corresponding K\"all\'en-Lehmann spectral density is:
\begin{eqnarray}
\rho(s) = &\theta(m^2-s)\,\, \frac{m^2-s_{peak}}{m^2+\mu_{lhd}^2-s_{peak}}
\,\,\delta(s-s_{peak}) \nonumber\\
+& \theta(s-m^2)\,\,\frac{\mu_{lhd}^2}
{[\,s-M^2-\mu_{lhd}^2\,\log((s-m^2)/\mu_{lhd}^2)\,]^2+\pi^2\,\mu_{lhd}^4} .
\end{eqnarray}

With these formulas we now try to describe the LEP data.
We start with the case $d=5$.
The delta-peak will be assumed to correspond to the
peak at 98 GeV, with a fixed value of $\alpha_{98}$.
Ultimately we will vary the location of the peak between
$95\gev < m_{peak} < 101\gev$ and $0.056 < \alpha_{98} < 0.144$.
After fixing $\alpha_{98}$ and $m_{peak}$ we have one free variable,
 which we take to be $\mu_{lhd}$. If we also take a fixed
value for $\mu_{lhd}$ all parameters and thereby the
spectral density is known. We can then numerically integrate the
spectral density over selected ranges of $s$. The allowed range of $\mu_{lhd}$ 
is subsequently determined by the data at 115 GeV.
Since the peak at 115 GeV is not very well constrained, we
demand here only that the integrated spectral density
from $s_{down} = (110\gev)^2$ to $s_{up} = (120\gev)^2$
is larger than 30\%. This condition, together with formula (15),
which implies:
\beq \rho(s) < \frac{(s-m^2)^{1/2}}{\pi\,\mu_{lhd}^3} , \eeq

leads to the important analytical result:
\beq 
\frac{2}{3\pi\,\mu_{lhd}^3} [\,(s_{up}-m_{peak}^2)^{3/2} - 
(s_{down}-m_{peak}^2)^{3/2}\,]
>0.3 \eeq
This implies $\mu_{lhd} < 53\gev$. Using the constraint from
the strength of the delta-peak, it follows that the continuum starts
very close to the peak, the difference being less than 2.5 GeV.
This allows for a natural explanation, why the CL for the fit in the
whole range from 100 GeV to 110 GeV is somewhat less than what is expected by
pure background. The enhancement can be due to a slight, spread-out Higgs signal.
Actually when fitting the data with the above conditions, one finds for small 
values of $\mu_{lhd}$, that the integrated spectral density in the range
100 GeV to 110 GeV can become rather large, which would lead to problems
with the 95\% CL limits in this range. We therefore additionally demand
that the integrated spectral density in this range is less than 30\%.
There is no problem fitting the data with these conditions. As allowed ranges
we find:
\begin{eqnarray}
& 95\gev< m < 101\gev \nonumber\\
& 111\gev< M < 121\gev \nonumber\\
& 26\gev < \mu_{lhd} < 49\gev 
\end{eqnarray}

As a final consistency check we notice that the results are in 
agreement with the upper limit on the Higgs boson mass from 
precision measurements $m_H~<~190\gev$. The integrated spectral density
above $190\gev$ is less than 1\%. This upper limit on the Higgs mass
is actually so large, because of the measurement of the b-asymmetry,
which is away from the other measurements, that are very close
together. One can follow the analysis from \cite{andrea} with  the latest
data and finds at the 95\% CL $m_H < 124\gev$ from the leptonic data
on $\mbox{sin}^2_\theta(lept)$, $m_H < 144\gev$ from $m_W$ and $m_H < 109\gev$
from the two combined. Even this limit is not inconsistent
with our model, which may be taken as additional confirmation.
One can actually allow for the integrated spectral density in the
low mass region to be larger than
 given in our analysis, if one allows for a fraction of invisible
decays of the Higgs boson. Indeed if this strong limit is valid,
a low mass spread-out Higgs boson appears to be the only possibility
consistent with experiment.

We now repeat the analysis for the case $d=6$.
The analytic argument gives the result:
\beq
\frac{s_{up}-s_{down}}{\pi^2\,\mu_{lhd}^2} > 0.3
\eeq
which implies $\mu_{lhd}<28\gev$.
Because of this low value of $\mu_{lhd}$ it is difficult
to get enough spectral weight arond 115 GeV and one also tends to get
too much density below 110 GeV. So the fit was only possible in a restricted range.
We found the following limits:

\begin{eqnarray}
& 95\gev< m < 101\gev \nonumber\\
& 106\gev < M < 111\gev \nonumber\\
& 22\gev < \mu_{lhd} < 27\gev 
\end{eqnarray}
As far as the limit from the precision data there is no problem here
either. The spectral weight above 190 GeV is a bit larger, because
the six-dimensional spectrum falls off somewhat more slowly at high values
of $s$, however is still less than 4\%. Though not quite ruled out, the six-dimensional
case therefore seems to be somewhat disfavoured compared to the five-dimensional
case.\\

From the discussion  it is clear that the data can be well described
by the assumption of a higher dimensional Higgs field. The model is renormalizable
and contains only two parameters beyond the SM. The relevant mass scales are
all comparable to each other and no fine tuning is present. The model obviously
fits the data better than pure background or the MSSM, due to the fact that
fractional Higgses are possible at 98 GeV and 115 GeV. Also it provides an
explanation for the slight discrepancy in the range 100 GeV to 110 GeV, that is not
explained by another model. What is not clear is how significant the overall results
are. For this an analysis of the confidence level
on a bin by bin basis would be necessary. From the discussion
given in \cite{lephiggs} we cannot reliably extract this significance.
This can only be done by the experimental groups. Assuming independence in the
different mass ranges a very rough estimate gives a $3.5\,\sigma - 4\,\sigma$
overall signal.\\

If this model is indeed true, what are the consequences for the Higgs search
at the LHC? The situation appears to be essentially hopeless. Most of the
spectral density is tied up in the low Higgs mass region. In this region
the Higgs search is performed at the LHC in the rare decay modes 
$H\rightarrow \gamma \gamma$ and $H \rightarrow Z Z^* \rightarrow l^+l^-l^+l^-$,
both of which heavily rely on an excellent mass resolution in order to extract
the Higgs signal. Since now there is no mass peak, these signals do not work.
In the range 150 GeV-180 GeV one can try to use the dominant $H\rightarrow W^*W^*$ decay
mode. However this does not work either, because the integrated spectral density
in this region is less than 2\% for the $d=5$ case and less than 6\%
for the $d=6$ case. Also the golden decay modes above the 
Z threshold are too small, since the integrated spectral density above 190 GeV is less
than 4\%.  The only possibility to study this type of Higgs boson appears to be
the international linear collider (ILC), where the Higgs signal can be studied
independently of its decay modes. The spectral density can be reconstructed from
the Bjorken process $e^+e^-\rightarrow Z H$, by looking at the invariant mass
of the Higgs boson, that can be extracted, if one knows the Z momentum precisely.
Since backgrounds and signal can be calculated and measured very precisely, this 
model presents no particular problem for the ILC.\\

{\bf Acknowledgement} We thank Dr. D. Schlatter for an illuminating
conversation. We thank Prof. M. Veltman,
 Drs. A. Ferroglia, A. Lorca and B. Tausk
 for discussions and careful readings of the manuscript.

\end{document}